\documentclass[twocolumn,showpacs,preprintnumbers,amsmath,amssymb, prb]{revtex4}
\usepackage{graphicx}
\usepackage{dcolumn}
\usepackage{bm}
\usepackage{wasysym}
\usepackage{textcomp}
\usepackage{epstopdf}
\begin{document}

\preprint{APS/123-QED}

\title{Field induced changes across magnetic compensation in Pr$_{1-x}$Gd$_x$Al$_2$ alloys}
\author{P. D. Kulkarni, A. K. Nigam, S. Ramakrishnan and A. K. Grover}
\affiliation{Department of Condensed Matter Physics and Materials
Science, Tata Institute of Fundamental Research, Homi Bhabha Road,
Colaba, Mumbai 400005, India.}
\date{\today}

\begin{abstract}
The magnetic compensation phenomenon has been explored in the Pr$_{1-x}$Gd$_x$Al$_2$ series. The contributions from Pr and Gd moments compensate each other at a specific temperature in the ordered state (below $T_c$). At high fields, the magnetic reorientation (with respect to the external field direction) of the Pr and Gd moments appears as a minimum in the thermomagnetic response. We demonstrate several interesting attributes related with the magnetic reorienation phenomenon, viz., oscillatory behavior of the magneto-resistance, sign change of the anamalous Hall resistivity, fingerprints of field induced changes in the specific heat and ac-susceptibility data.  
\end{abstract}



\maketitle

Historically, the magnetic compensation behaviour in the admixed rare earth intermetallics was first reported by Williams $et$~$al.$ \cite{Williams}. They had shown that the compensation points exist in Pr$_{1-x}$Gd$_x$Al$_2$ ($x$ = 0.2-0.3) and the substitution of upto 20 atomic \% of Pr by Gd results in lower magnetization values compared to the pure alloy. We have revisited the Pr$_{1-x}$Gd$_x$Al$_2$ series in the light of the results in recent years in Sm$_{1-x}$Gd$_x$Al$_2$ \cite{Adachi1, Adachi2, Adachi3, Adachi4, Taylor, Chen, Wu, Qiao} pertaining to the magnetic compensation phenomenon \cite{Grover1, Grover2, Grover3} and investigated the field-induced changes across $T_{comp}$\cite{Rakhecha, Kulkarni}. The Samarium ion, due to the admixture of the higher order multiplets in its ground state in the presence of the crystalline electric field (CEF), has different temperature dependences for its spin and orbital parts \cite{Adachi3, Buschow, Malik1, Malik2}. The orbital and spin contributions to the total moment of the Samarium ion are comparable and coupled antiparallel in the ground multiplet level as a consequence of the spin-orbit interaction, leaving a small `orbital-surplus' magnetization ($\sim$ 0.71 $\mu_{\rm B}$/ion) with the free Sm$^{3+}$ ion. However, in a metallic matrix, the conduction electron contribution could aid the spin part to drive the Samarium metal to a 'spin-surplus' system and reduce the difference moment down to $\sim$ 0.1 $\mu_{\rm B}$ per Sm$^{3+}$ ion\cite{Adachi2}. Samarium metal can be combined with different non-magnetic elements to form the `orbital' or `spin' surplus alloys \cite{Adachi1, Adachi2}. Adachi and Ino re-kindled the interest\cite{Adachi1} in magnetic compensation behaviour in the Samarium based alloys by their observation that the large spin polarization co-existing with no bulk magnetization could have niche applications. It has been shown by Chen $et$ $al.$ \cite{Chen} in Sm$_{1-x}$Gd$_x$Al$_2$ ($x$ = 0.01 and 0.02) that the field-induced reversal in the orientations of the magnetic moments of Sm and Gd in the Sm$_{1-x}$Gd$_x$Al$_2$ ($x$ = 0.01 and 0.02) alloys imprints as a sharp peak centred around a given $T_{comp}$ value in the heat capacity data. The observation that the peak height scales with the applied field in $x$ = 0.01 alloy could imply the occurence of an entropic change and a phase trasition like response at $T_{comp}$. They also reported that the usual negative magnetoresistivity below the magnetic ordering temperature reverses its sign across $T_{comp}$ \cite{Chen}. 
Our new findings in Pr$_{1-x}$Gd$_x$Al$_2$ series being reported here emphasize the generic nature of the field-induced pseudo-phase transition in the admixed rare earth systems showing magnetic compensation behaviour. The results in Sm based alloys are therefore not unique to special characteristics of Sm$^{3+}$ ions under the influence of CEF and exchange field effects. The sign change in the magneto-resistance across $T_{comp}$ is seen in Pr$_{0.8}$Gd$_{0.2}$Al$_2$ alloy as well, however, it is additionally accompanied by an oscillatory behaviour at lower temperatures. A further interesting result in Pr$_{0.8}$Gd$_{0.2}$Al$_2$ is the identification of the fingerprint of the magnetic reorientation in the in-field ac-susceptibility data. To fortify the field-induced changes across the magnetic compensation temperature, we are also presenting the results in Pr$_{0.83}$Gd$_{0.17}$Al$_2$, where the $T_{comp}$ and $T_c$ are in close proximity to each other.  

A series of polycrystalline Pr$_{1-x}$Gd$_x$Al$_2$ ($x$ = 0, 0.15, 0.17, 0.2 and 0.25) alloys were prepared by melting together the stoichiometric amounts of the constituent elements in a tetra arc furnace (Model: TCA 4-5, Techno Search Corp., Japan). The elemental analysis of the admixed alloys using an analyzer (JEOL JSX-3222) reassured the targeted stoichiometries. The x-ray diffraction patterns were recorded for the powdered samples using X'pert PRO x-ray diffractometer. Indexing of the $x$-ray patterns confirmed the cubic $C$15 phase in all the alloys. 
\begin{figure}[h]
\includegraphics[width=0.45\textwidth]{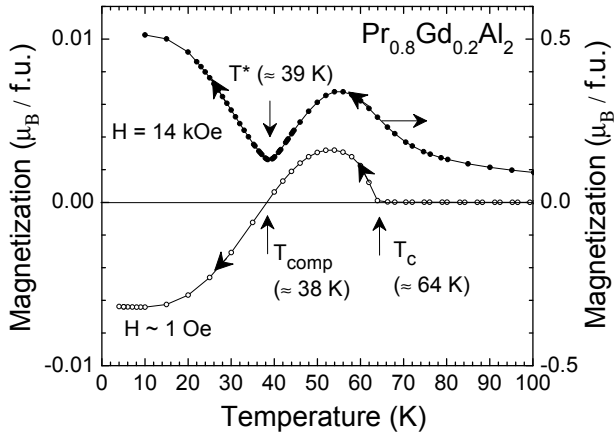}
\caption{\label{fig1} Field cooled cooldown magnetization ($M_{FCC}$) in Pr$_{0.8}$Gd$_{0.2}$Al$_2$ alloy. The $T_c$ ($\HF$ 64 K) and $T_{comp}$ ($\HF$ 38 K) are marked in the nominal zero field cooled curve. In high field (14 kOe), the occurrence of magnetic turnaround results in a minimum at $T$* $\HF$ 39 K.}
\end{figure}
A small piece of Pr$_{0.83}$Gd$_{0.17}$Al$_2$ was annealed at 1000~{\textcelsius} for 10 days to ascertain the differences in the results in the as-grown and the annealed samples.  The dc magnetization and the ac susceptibility data were recorded using Quantum Design (QD) Inc. superconducting quantum interference device (SQUID) magnetometer (Model MPMS-5). The heat capacity and the resistivity data was measured in a Physical Property Measurement System (PPMS) of QD Inc. U. S. A. The Hall resistance is measured
as a function of temperature using the homemade setup for transport studies.

In Fig. 1, the field cooled cooldown (FCC) magnetization ($M_{FCC}$) curves in Pr$_{0.8}$Gd$_{0.2}$Al$_2$ are shown in $H$ $\sim$ 1 Oe and $H$ = 14 kOe. In the nominal zero field ( $H$ $\sim$ 1 Oe) cooled data, the magnetic ordering temperature of this alloy is marked ($T_c$ $\HF$ 64 K). The magnetization signal is positive between $T_c$ and $T_{comp}$ ($\HF$ 38 K) and negative below $T_{comp}$. At high fields, a minimum in the thermomagnetic curve is observed at $T^*$ $\HF$ 39 K.  

\begin{figure}[h]
\includegraphics[width=0.45\textwidth]{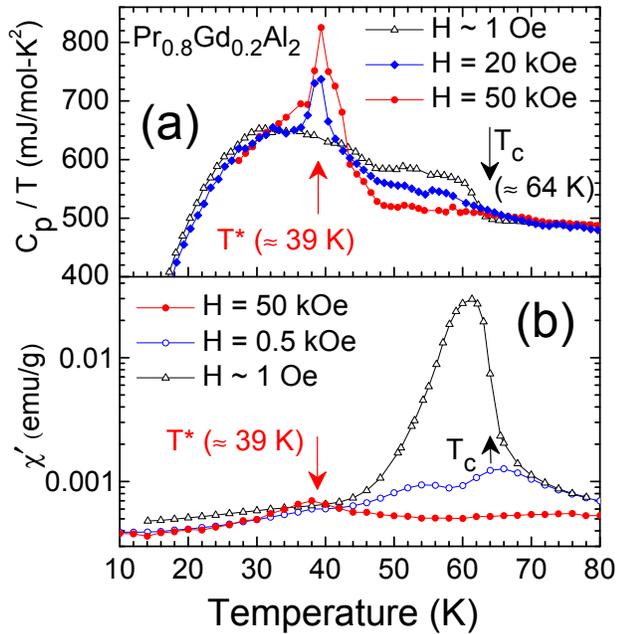}
\caption{\label{fig2}(Color online) (a) Specific heat ($C_p$/$T$ versus $T$) data in $H$ = 1 Oe, 20 kOe and 50 kOe and (b) the temperature dependence of the ac-susceptibility in $H$ = 1 Oe, 0.5 kOe and 50 kOe in Pr$_{0.8}$Gd$_{0.2}$Al$_2$ alloy. In panel (b), $T$* ($\approx$ 39 K) in $H$ = 50 kOe is marked.}
\end{figure}

Figure 2 shows a collation of the temperature dependences of (a) the specific heat in $H$ = 1 Oe, 20 kOe and 50 kOe and (b) the ac susceptibility responses in $H$ = 1 Oe, 500 Oe, 50 kOe in Pr$_{0.8}$Gd$_{0.2}$Al$_2$ alloy. In Fig. 2 (a), the nominal zero field specific heat data shows a rise in the specific heat starting at $T_c$ ($\HF$ 64 K), followed by a broad hump. The broad peak closer to $T_c$ gets suppressed as the magnetic field is progressively enhanced, and a relatively sharp field-induced peak develops at $T$*, whose height scales with the applied magnetic field (all data not shown). In the ac-susceptibility data in the nominal zero field (cf. Fig. 2(b)), a broad peak can be seen at the magnetic transition ($T_c$ is marked at the rising edge of the peak). At high fields, this gets collapsed and an additional peak at lower temperature surfaces up at $T$*.  
 
\begin{figure}[h]
\includegraphics[width=0.45\textwidth]{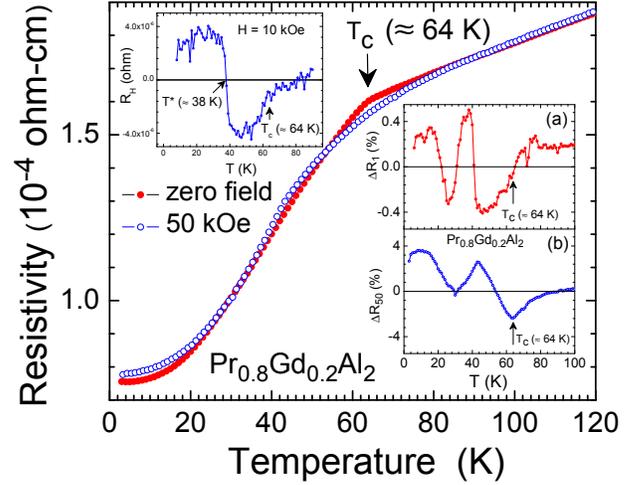}
\caption{\label{fig3}(Color online) The electrical resistivity as a function of $T$ in Pr$_{0.8}$Gd$_{0.2}$Al$_2$ alloy in nominal zero field and 50 kOe. An inset in the top left corner shows the Hall resistance ($R_H$) as a function of $T$ in H = 10 kOe. Insets, (a) and (b) in bottom right show the normalized magnetoresistance ($\vartriangle$R(H) = (R(H) - R(0))/R(0)) in (a) 1 kOe and (b) 50 kOe.}
\end{figure}

Figure 3 shows the electrical resistivity as a function of temperature in Pr$_{0.8}$Gd$_{0.2}$Al$_2$ alloy. The nominal zero field and 50 kOe data are plotted together for comparison. The sharp drop in the nominal zero field electrical resistance data occurs at $T_c$ ($\approx$ 64 K). This feature broadens out in 50 kOe and lies below the $R$($T$) curve in the nominal zero field. The inset panels (a) and (b) in Fig. 3 show the magnetoresistance values calculated from the electrical resistance data recorded in 1 kOe and 50 kOe. The normalized magnetoresistance in 1 kOe ($\triangle$$R_{1}$ = [R(1 kOe) - R(0)]/R(0)) displays negative values below $T_c$ and then crosses over to the positive values at $\approx$ 40 K. Below this temperature, the magnetoresistance changes sign two more times exhibiting an oscillatory character. The percent change in magnetoresistance in 1 kOe is within 0.5 \%. In 50 kOe, the magnetoresistance retains the oscillatory variation and it is an order of magnitude higher. Inset in the top left corner of Fig. 3 shows the Hall resistance as a function of temperature in H = 10 kOe. The sign change in Hall resistance ($R_H$) can be observed at 38 K.  

Figure 4(a) shows the $M_{FCC}$ in nominal zero field in Pr$_{0.83}$Gd$_{0.17}$Al$_2$ and Pr$_{0.75}$Gd$_{0.25}$Al$_2$. The $T_c$ values are marked at $\HF$ 55 K and $\HF$ 72 K, respectively. Both the alloys display the magnetic compensation behavior, their $T_{comp}$ values are marked at 49 K and 33 K, respectively. Note that the $T_{comp}$ and $T_c$ are well separated in Pr$_{0.75}$Gd$_{0.25}$Al$_2$, while in Pr$_{0.83}$Gd$_{0.17}$Al$_2$ these differ only by 6 K. An inset in Fig. 4 (a) shows  $M_{FCC}$($T$) response in Pr$_{0.85}$Gd$_{0.15}$Al$_2$ at $H$ = 55 and 100 Oe. The zero crossover in lower field of 50 Oe and turnaround in 100 Oe can be seen at 50 K, $T_c$ of the sample is marked at 53 K.   

\begin{figure}[h]
\includegraphics[width=0.45\textwidth]{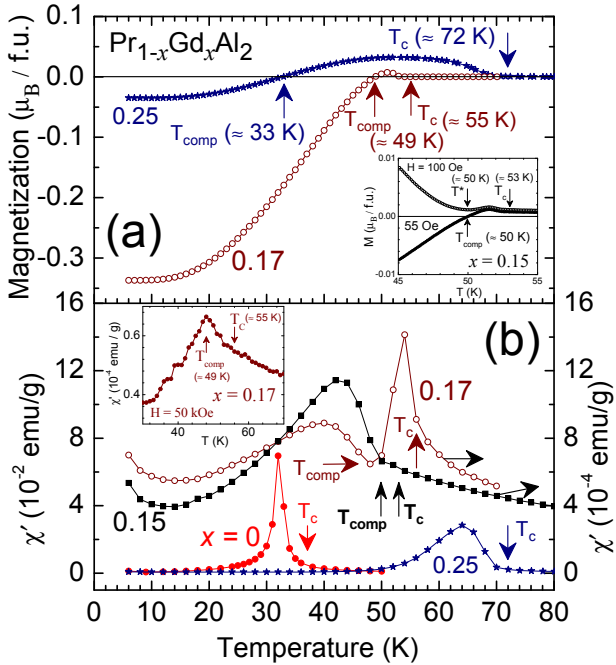}
\caption{\label{fig4}(Color online) (a) The $M_{FCC}$($T$) in nominal zero field in Pr$_{0.83}$Gd$_{0.17}$Al$_2$ and  Pr$_{0.75}$Gd$_{0.25}$Al$_2$ alloys. The $T_{comp}$ and $T_c$ for both the alloys are marked. An inset in Fig. 4 (a) shows the $M_{FCC}$($T$) in Pr$_{0.85}$Gd$_{0.15}$Al$_2$ in 55 Oe and 100 Oe. $T_{comp}$ and $T_c$ values in Pr$_{0.85}$Gd$_{0.15}$Al$_2$ are 50 K and 53 K, respectively. (b) The nominal zero field cooled in-phase ac-susceptibility in alloys with $x$ = 0, 0.15, 0.17, and 0.25. Note the two peak ac-response in the Pr$_{0.83}$Gd$_{0.17}$Al$_2$. An inset in panel (b) shows the ac-susceptibility peak in 50 kOe (the corresponding $T_{comp}$ is marked) in Pr$_{0.83}$Gd$_{0.17}$Al$_2$.}
\end{figure}

Figure 4(b) shows the ac-susceptibility responses in Pr$_{1-x}$Gd$_x$Al$_2$ ($x$ = 0, 0.15, 0.17 and 0.25) series. In pure PrAl$_2$, the ac-signal in nominal zero field starts rising at $\sim$ 37 K, which corresponds to its $T_c$\cite{Williams}. The substitution of 25 atomic \% Gd$^{3+}$ in PrAl$_2$ shifts the onset of ac-peak to $\sim$ 72 K. The ac-responses in 15 \% and 17 \% doped samples are two orders lower in magnitude and are therefore shown with respect to the scale on the right-hand side in Fig. 4(b). The ac-response in 17 \% doped sample has two peaks, a sharp peak above $T_{comp}$ (closer to its $T_c$) and a broad peak below $T_{comp}$. In Pr$_{0.85}$Gd$_{0.15}$Al$_2$, the broad peak has larger magnitude compared to the corresponding peak in Pr$_{0.83}$Gd$_{0.17}$Al$_2$. The peak above $T_{comp}$ in this case appear as an added tail to the broad peak at lower temperatures. Note that the $T_{comp}$ and $T_c$ are separated by $\sim$ 3 K in this alloy. An inset in Fig. 4 (b) displays the in-field ($H$ = 50 kOe) ac-susceptibility peak in Pr$_{0.83}$Gd$_{0.17}$Al$_2$ alloy at $\HF$ 48 K. It matches with the turnaround temperature $T$* of the thermomagnetic curve in 50 kOe in this alloy. (data not shown here). 

\begin{figure}[h]
\includegraphics[width=0.45\textwidth]{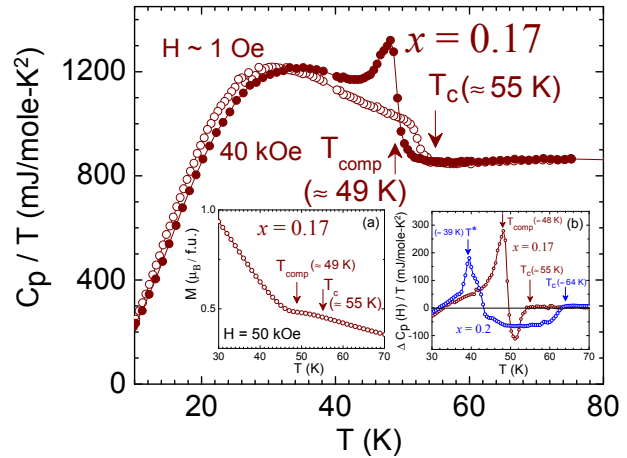}
\caption{\label{fig5}(Color online) (a) The temperature dependence of the specific heat in  nominal zero field and 40 kOe in Pr$_{0.83}$Gd$_{0.17}$Al$_2$. Inset (a) in Fig. 5 shows a portion of $M_{FCC}$($T$) in $x$ = 0.17 in $H$ = 50 kOe. Inset (b) shows the \textquoteleft difference specific heat\textquoteright, $\triangle$$C_p(H)$/$T$ (= $C_p(H)$/$T$ - $C_p(0)$/$T$), in Pr$_{0.83}$Gd$_{0.17}$Al$_2$ and Pr$_{0.8}$Gd$_{0.2}$Al$_2$.}
\end{figure}

In the Pr$_{1-x}$Gd$_x$Al$_2$ series with $x$ = 0.15, 0.17, 0.2 and 0.25, the contribution to magnetization signals of the Pr$^{3+}$ and Gd$^{3+}$ moments are phase reversed and get compensated at specific temperatures ($T_{comp}$) below $T_c$. A comparison of the responses in $H$ $\HF$ 10 kOe in this series shows that the magnetization in Pr$_{0.8}$Gd$_{0.2}$Al$_2$ remains closest to the zero value at 5 K. The increase in the doping concentration of Gd$^{3+}$ ions in PrAl$_{2}$ increases $T_{c}$ and decreases $T_{comp}$ values in the admixed series.
At high fields, the magnetic compensation between the Pr$^{3+}$ and Gd$^{3+}$ manifests as a minimum at $T^*$ due to reorientation of these antiferromagnetically coupled entities. We have probed the magnetic reorientation process using the ac susceptibility measurements at high fields. At $H$ = 50 kOe in Pr$_{0.8}$Gd$_{0.2}$Al$_2$ (Fig. 2(b)), the ac peak emerges close to the $T$* and can be identified with the reorientation of Pr and Gd moments with respect to the external field direction. In Pr$_{0.83}$Gd$_{0.17}$Al$_2$ the presence of the external magnetic field restricts the rare earth moments to respond to ac-fields and the two-peak structure in the ac-susceptibility shown in Fig. 4(b) disappears. However, the magnetic reorientation driven by the realignment of the rare earth moments produces ac-response and an additional peak emerges centred around $T$* (see the inset in Fig. 4(b)). 

The positions of the peaks in the ac-responses (cf. Fig. 4) show the effect of the Gd$^{3+}$ substitution on the magnetic ordering process. The magnitude of the ac-susceptibility peak reduces by two orders with the 15 \% substitution of Gd$^{3+}$ ions in pure PrAl$_2$. A possible mechanism could be the reduced magnetization of the domains because of the antiferromagnetic coupling between the Gd$^{3+}$ and Pr$^{3+}$ moments. Pr$_{0.85}$Gd$_{0.15}$Al$_2$ alloy appears to be on the verge of nucleating domains which has the dominance of Gd$^{3+}$ moments (note the shoulder before the rising ac-peak in Fig. 4(b)), while a sharp peak near the same position has emerged in the ac-response in Pr$_{0.83}$Gd$_{0.17}$Al$_2$. The sharp ac-peak in Pr$_{0.83}$Gd$_{0.17}$Al$_2$ could be attributed to the freezing in of the domains in which magnetization from Gd$^{3+}$ ions dominates close to the magnetic ordering temperature, followed by the ac-response of all the domains realigning during the magnetic compensation process. With 20-25 atomic \% substitution of the Gd$^{3+}$ ions in PrAl$_2$, the ac-response is completely dominated by the Gd moments and the dynamics of all the antiferromagnetically coupled Pr$^{3+}$ momemts also slows down right at the $T_c$. 

The magnetic reorientation process also leaves an imprint in the temperature dependence of the specific heat in Pr$_{0.8}$Gd$_{0.2}$Al$_2$. In nominal zero field (Fig. 2(a)) the magnetic transition at $T_c$ is captured, however, at high fields, an additional peak surfaces up at $T*$ $\HF$ 39 K. The emergence of the specific heat peak at $T$* indicates that the magnetic reorientation has an attribute of the pseudo-phase transition in these alloys. 

The temperature dependence of the specific heat in Pr$_{0.83}$Gd$_{0.17}$Al$_2$ is shown in the Fig. 5. Inset (a) in Fig. 5 shows a portion of the thermomagnetic curve at high fields (50 kOe) which does not have the usual minimum close to $T_{comp}$. The peak in the \textquoteleft difference specific heat\textquoteright\ data in $H$ = 20 kOe close to $T_c$ in Pr$_{0.83}$Gd$_{0.17}$Al$_2$ appears to be sharp (see inset (b) in Fig. 5). This sharpness can be compared with the peak in its high field ac-susceptibility data (see inset in Fig. 4(b)). These two sharp features support the notion of field induced phase-transition at compensation temperature, as advocated by Chen $et$~$al.$ \cite{Chen} in the Sm$_{0.98}$Gd$_{0.02}$Al$_2$ alloy, where the magnetic compensation phenomenon was attributed to the special properties of the Sm$^{3+}$ ions \cite{Adachi2, Adachi3}. Chen $et$~$al.$ \cite{Chen} also observed the sign change (-ve to +ve) in the magnetoresistance across $T_{comp}$ in Sm$_{0.98}$Gd$_{0.02}$Al$_2$ alloy. We do not observe this correlation in Pr$_{1-x}$Gd$_x$Al$_2$ series. However, the sign change in the Hall resistance correlates to the results in Sm$_{0.98}$Gd$_{0.02}$Al$_2$ alloy. The spin-disorder resistivity which freezes at $T_c$, again appears to get alive while cooling the sample in the presence of the high external magnetic field. This can occur if the spin-orbit configuration continues to undergo a rearrangement, which is the case in these alloys during in-field cooling. We believe that the oscillatory nature of the magnetoresistance (see Insets (a) and (b) in Fig. 3) is a generic feature in the admixed rare earth intermetallics showing the compensation behavior. It is fruitful to recall here that oscillatory magneto-resistance response also stands reported in a single crystal of Nd$_{0.75}$Ho$_{0.25}$Al$_2$\cite{Kulkarni1}.  

To summarize, the magnetic compensation behaviour in the  Pr$_{1-x}$Gd$_x$Al$_2$ series has been studied in the contemporory context. The fingerprint of magnetic turnaround across $T_{comp}$ is identified in the in-field ac-susceptibility data. The temperature dependence of the in-field specific heat supports the notion of field-induced phase transition across the $T_{comp}$ and it corroborates the earlier observation in the single crystal Nd$_{0.75}$Ho$_{0.25}$Al$_2$ \cite{Kulkarni1}. A curius oscillatory behavior of the magnetoresistance as a function of temperature is observed in the Pr$_{0.8}$Gd$_{0.2}$Al$_2$ alloy. The change in sign of the Hall voltage across $T_{comp}$ is also an important observation. It should be of interest to explore the temperature dependences of the 4$f$-spin and 4$f$-orbital contributions of Pr$^{3+}$ and 4$f$-spin contribution of Gd$^{3+}$ via x-ray magnetic circular dichroism measurements in single crystals of Pr$_{1-x}$Gd$_x$Al$_2$ alloys.

We thank S. K. Dhar and P. L. Paulose for their association in the initial phase of this work. We also acknowledge D. D. Buddhikot for his help in some of the measurements.

\end{document}